 \documentclass[twocolumn,twoside,showpacs,superscriptaddress,amsmath,amssymb,prl]{revtex4}
 
 \usepackage{dcolumn}
\usepackage{cases}
 \usepackage{CJKutf8}
 \usepackage{txfonts}
 \usepackage{mathrsfs}
 \usepackage{amssymb}
 \usepackage[dvips]{graphicx}
 \usepackage{bm}
 \usepackage{subfigure}
 \usepackage{floatflt}
 \usepackage{float}
 \usepackage{stfloats}
 \usepackage{rotating}
 \usepackage{multirow}
 \usepackage[bookmarksnumbered,bookmarksopen,colorlinks,citecolor=blue,linkcolor=blue]{hyperref} %
 \usepackage{booktabs}
 \usepackage{epsfig,graphics,psfrag,amsmath}
 \usepackage{sidecap}
   \usepackage{ulem}

\def\thu{Department of Physics, Tsinghua University, Beijing 100084, China}
\def\nju{School of Physics, Nanjing University, Nanjing 210093, China}

\def\imp{Institute of Modern Physics, Chinese Academy of Sciences, Lanzhou 730000, China}
\def\sinap{Shanghai Institute of Applied Physics, Chinese Academy of Science, Shanghai 201800, China}
\def\sari{Shanghai Advanced Research Institute, Chinese Academy of Science, Shanghai 201210, China}
\def\cicq{Collaborative Innovation Center of Quantum Matter, Tsinghua University, Beijing 100084, China}

\def\buaa{School of Physics, Beihang University, Beijing 102206, China}
\def\Wigner { Wigner Research Centre for Physics, Budapest 1121, Hungary}
\def\INR {Institute for Nuclear Research, National Academy of Sciences of Ukraine, Kyiv 03680, Ukraine}

\def\krsn{$^{86}$Kr+$^{124}$Sn}

\newbox\ftyuebox\newdimen\ftyuewd\font\ftyuea=ftyuea at 57.816truept\setbox\ftyuebox=\vbox{\hbox{\ftyuea\char0}}\ftyuewd=\wd\ftyuebox\setbox\ftyuebox=\hbox{\vbox{\hsize=\ftyuewd\parskip=0pt\offinterlineskip\parindent0pt\hbox{\ftyuea\char0}}}\ifx\parbox\undefined\def\setftyue{\box\ftyuebox}\else\def\setftyue{\parbox{\wd\ftyuebox}{\box\ftyuebox}}\fi
\DeclareUnicodeCharacter{73A5}{\raise0.2ex\hbox{\setftyue}}

\begin{document}

\title{Probing high-momentum component in nucleon momentum distribution by neutron-proton bremsstrahlung $\gamma$-rays in heavy ion reactions}
\date{\today}

\author{Yuhao Qin \begin{CJK*}{UTF8}{gbsn}(秦雨浩)\end{CJK*}} 
\affiliation{\thu}
 
\author{Qinglin Niu \begin{CJK*}{UTF8}{gbsn}(牛庆林)\end{CJK*}} 
\affiliation{\nju}

\author{Dong Guo \begin{CJK*}{UTF8}{gbsn}(郭栋)\end{CJK*}} 
 \affiliation{\thu}

\author{Sheng Xiao \begin{CJK*}{UTF8}{gbsn}(肖圣)\end{CJK*}}  
\affiliation{\thu}

\author{Baiting Tian \begin{CJK*}{UTF8}{gbsn}(田柏汀)\end{CJK*}} 
\affiliation{\thu}

\author{Yijie Wang \begin{CJK*}{UTF8}{gbsn}(王轶杰)\end{CJK*}} 
\affiliation{\thu}

 \author{Zhi Qin \begin{CJK*}{UTF8}{gbsn}(秦智)\end{CJK*}} 
 \affiliation{\thu}

\author{Xinyue Diao \begin{CJK*}{UTF8}{gbsn}(刁昕玥)\end{CJK*}} 
\affiliation{\thu}
 
\author{Fenhai Guan \begin{CJK*}{UTF8}{gbsn}(关分海)\end{CJK*}} 
\affiliation{\thu}

 \author{Dawei Si \begin{CJK*}{UTF8}{gbsn}(司大伟)\end{CJK*}}    
\affiliation{\thu}

\author{Boyuan Zhang \begin{CJK*}{UTF8}{gbsn}(张博源)\end{CJK*}}  
\affiliation{\thu}

\author{Yaopeng Zhang \begin{CJK*}{UTF8}{gbsn}(张耀鹏)\end{CJK*}}  
\affiliation{\thu}

\author{Xianglun Wei \begin{CJK*}{UTF8}{gbsn}(魏向伦)\end{CJK*}} 
\affiliation{ \imp }

\author{Herun Yang \begin{CJK*}{UTF8}{gbsn}(杨贺润)\end{CJK*}} 
\affiliation{ \imp }

\author{Peng Ma \begin{CJK*}{UTF8}{gbsn}(马朋)\end{CJK*}} 
\affiliation{ \imp }

\author{Rongjiang Hu \begin{CJK*}{UTF8}{gbsn}(胡荣江)\end{CJK*}} 
\affiliation{ \imp }

\author{Limin Duan \begin{CJK*}{UTF8}{gbsn}(段利敏)\end{CJK*}} 
\affiliation{ \imp }

\author{Fangfang Duan \begin{CJK*}{UTF8}{gbsn}(段芳芳)\end{CJK*}} 
\affiliation{ \imp }

\author{Qiang Hu \begin{CJK*}{UTF8}{gbsn}(胡强)\end{CJK*}} 
\affiliation{ \imp }

\author{Junbing Ma \begin{CJK*}{UTF8}{gbsn}(马军兵)\end{CJK*}} 
\affiliation{ \imp }

\author{Shiwei Xu \begin{CJK*}{UTF8}{gbsn}(许世伟)\end{CJK*}} 
\affiliation{ \imp }

\author{Zhen Bai \begin{CJK*}{UTF8}{gbsn}(白真)\end{CJK*}} 
\affiliation{ \imp }

\author{Yanyun Yang \begin{CJK*}{UTF8}{gbsn}(杨彦云)\end{CJK*}} 
\affiliation{ \imp }

\author{Hongwei Wang \begin{CJK*}{UTF8}{gbsn}(王宏伟)\end{CJK*}} 
\affiliation{\sinap}
\affiliation{\sari}

\author{Baohua Sun \begin{CJK*}{UTF8}{gbsn}(孙保华)\end{CJK*}} 
\affiliation{ \buaa }

\author{Sergei P. Maydanyuk} 
\affiliation{\Wigner}
\affiliation{\INR}

\author{Chang Xu \begin{CJK*}{UTF8}{gbsn}(许昌)\end{CJK*}}   \email{cxu@nju.edu.cn}
\affiliation{\nju}

\author{Zhigang Xiao \begin{CJK*}{UTF8}{gbsn}(肖志刚)\end{CJK*}}  \email{xiaozg@tsinghua.edu.cn}
\affiliation{\thu}
 \affiliation{\cicq}
\begin{abstract}

    The high momentum tail (HMT) of nucleons, as a signature of the short-range correlations in nuclei, has been investigated by the high-energy bremsstrahlung $\gamma$ rays produced in \krsn~ at 25 MeV/u.  The energetic photons are measured by a CsI(Tl) hodoscope mounted on the spectrometer CSHINE. The energy spectrum above 30 MeV can be reproduced by the IBUU model calculations incorporating the photon production channel from $np$ process in which the HMTs of nucleons is considered. A non-zero HMT ratio of about $15\%$ is favored by the data. The effect of the capture channel $np \to d\gamma$ is demonstrated. 
    
\end{abstract}

\maketitle

{\it  Introduction.} Short-range correlations (SRCs) are induced by the existence of tensor force and repulsive core between pairs of strongly interacting nucleons at short distances, leading to high momentum tails (HMTs) in the nucleon momentum distributions beyond the Fermi surface \cite{rmp2017}. SRCs provide insight into not only the correlated quantum nuclear many-body system, but also the nature of the strong nuclear interaction and how they are generated from quarks in nucleons \cite{arnps2017,emc83}. 

Much of our present understanding of SRCs comes from high-energy electron scattering measurements off finite nuclei \cite{clas14,clas18,clas19}, and from proton induced reactions on heavy targets \cite{Tang2003,Pias2006} as well as knockout reactions in inverse kinematics \cite{Patsyuk2021}.  Experiments have shown that nucleons in finite nuclei can form SRC pairs with small center-of-mass (CM) momentum and large relative momentum where the neutron-proton($np$) pairs are almost 20 times as prevalent as the proton-proton pairs \cite{clas18}. State-of-art approaches have been performed for finite nuclei and demonstrated that the HMT of $np$-SRCs is a universal feature using Variational Monte Carlo calculations \cite{Schiavilla:2006xx}, Self-consistent Green's function method \cite{Rios:2013zqa}, Cluster expansion method \cite{Alvioli:2016wwp}, Tensor-optimized high-momentum approach \cite{Lyu2019} and so on. The high-momentum feature of SRCs may shed light on the nuclear equation of state of nuclear matter at supra-saturation density \cite{Xu2012}, which has important implications in determining the mass and structure of stellar objects such as neutron stars \cite{LI201829}.

Energetic photons from intermediate energy nucleus–nucleus collisions present a unique opportunity to study strongly interacting nuclear systems  \cite{Remington87}. Different bremsstrahlung mechanisms including coherent and incoherent processes have been proposed to account for the emission of photons from nucleus–nucleus collisions  \cite{Ko85,Bauer86,Jetter94,Wang2020}  as well as  proton-nucleus collisions \cite{PhysRevLett.88.122302,PhysRevC.91.024605}. The neutron-proton ($np$) bremsstrahlung in the early stage of collisions is considered to be a main source of energetic photons, which could be used to probe SRCs in finite nuclei heavier than carbon-12. Note that SRCs between $np$ pairs increase largely the average nucleon kinetic energies of both projectile and target \cite{Xu2012}, and thus they are predicted to increase the high-energy photon production as well \cite{xue16}. The merit of high-energy photons is that they interact with nucleons only electromagnetically, which could serve to more precisely establish the behavior of HMTs \cite{xue16,yong17,PhysRevC.104.034603}.  However, due to experimental difficulties, high-energy photon bremsstrahlung data from heavy ion collisions are far too sparse to provide the necessary information required in the analysis of SRCs. Large uncertainties exist also in the available basic $np\to np\gamma$ bremsstrahlung data  \cite{Malek91}, which are necessary information for transport simulations and bremsstrahlung models such as meson exchange approaches  \cite{Cassing1990} .

In this letter, we report a novel measurement of the bremsstrahlung $\gamma$ spectrum in the reactions of \krsn~ at 25 MeV/u by using a CsI(Tl) hodoscope. The $\gamma$ energy spectrum up to 80 MeV is obtained experimentally, and analyzed using the isospin-dependent Boltzmann-Uehling-Uhlenbeck (IBUU) simulations \cite{LI2008113}, in which the $np\to np\gamma$ production channel is modeled and the HMTs of nucleons are included with adjustable ratios. The theoretic calculations reproduce the high-energy part of the $\gamma$ spectrum fairly well, supporting the bremsstrahlung $\gamma$ as a new way to study the SRCs in medium-mass nuclei.       



{\it  Experimental Setup.}
The experiment was performed at the radioactive beam line at Lanzhou (RIBLL1) \cite{ribll03}. The beam of 25 MeV/u $^{86}$Kr was delivered by the heavy ion research facility at Lanzhou (HIRFL) and bombarded on the $^{124}$Sn target installed in the chamber of the compact spectrometer for heavy ion experiment (CSHINE) \cite{NST-CSHINE,NIMA-CSHINE}, located at the final focal plane of RIBLL1. Four silicon strip detector telescopes (SSDTs), covering the angular range of $20^\circ<\theta_{\rm lab}<100^\circ $ in partial azimuth, were installed to measure the light charged particles (LCPs) and the intermediate mass fragments (IMFs) produced in the reactions. Fission fragments (FFs) are
measured by three Parallel Plate Avalanche Counters (PPACs) installed at $40^\circ$, $90^\circ$ and $-40^\circ$ with respect to the beam, respectively. For the detector performance of the CSHINE, one can refer to  \cite{NIMA-CSHINE,NIMA-CSHINE-SSD}. 

The $\gamma$-rays were measured by the closely packed 15-unit CsI(Tl) hodoscope installed at $\theta_{\rm lab}=110^\circ$, with the distance-to-target  $L$=110 cm. The 15 units form a $4\times4$ configuration with one corner missing. Each CsI(Tl) unit has the size of $7\times 7 \times 25 ~{\rm cm^3}$. The hodoscope was calibrated using $\gamma$ source and the high energy $\gamma$ rays from ${\rm  ^{19}F(p,\alpha \gamma) ^{16}O}$ and  ${\rm ^{7}Li(p,\gamma) ^{8}Be}$ reactions, respectively. The energy resolution of the individual units follows  $1.6\%+2\%/E_{\gamma}^{1/2}$, while the linearity is better than $2\%$ as tested at 17.6 MeV. 
For the details of the  assembly, calibration, simulations, performances of the $\gamma$ hodoscope and the experimental details, one can refer to \cite{NIMA-CSHINE-Gamma}. 

In order to measure the bremsstrahlung photons produced in the collisions,  the high energy $\gamma$ rays in coincidence with 2 LCPs recorded by the SSDTs or with 2 FFs recorded by the PPACs are analyzed. According to transport model calculations, this coincidence condition selects the semi-central events. 

{\it  Analysis and model description.}
The total energy of the $\gamma$ ray is reconstructed by adding the energy deposits in all fired units correlated in a given time window. Meanwhile, the incident position of the $\gamma$ ray and the transverse spatial spread of the event can be calculated. To be more convincing that the whole energy is collected, the unit with the maximum energy deposit, called {\it fire center}, is limited in the central $2\times2$ units of the hodoscope.  With the algorithm {\it ad hoc} developed for the energy reconstruction, the detection efficiency is nearly constant at about $80\%$ for incident $\gamma$-rays with $E_{\gamma}>20$ MeV \cite{NIMA-CSHINE-Gamma}. 

Fig. 1 (a) presents the vertical spatial spread  $\delta_{\rm y}$ as a function of the reconstructed total energy $E_{\rm recon}$. Here $\delta_{\rm y}$ is defined by $\delta_{\rm y}=\sum{E_{\rm i} {|y_{\rm i}-\bar{y}|}/E_{\rm tot}}$, where $E_{\rm i}$ and $y_{\rm i}$ are the energy and vertical position of the unit $i$ being fired, respectively, and $\bar{y}$ is the weighted vertical center of the  $\gamma$ incidence.  By the definition, $\delta_{\rm y}=0$ is expected if only one unit is fired, while  $\delta_{\rm y}=3.5$ cm means that two neighbouring units are fired with approximately the same energy deposit.  It is shown that the events are separated in two groups by $E_{\rm recon}=80$ MeV.  For the events with  $E_{\rm recon}<80$ MeV, the majority is contributed by  either a single fired CsI(Tl) unit or two neighbouring units, characterized by $\delta_{\rm y}=0$ or $0<\delta_{\rm y}<3.5$ cm, respectively. Some  events below 20 MeV are populated with $\delta_{\rm y}>3.5$ cm, suggesting that more than 2 units are fired. Given such low energy deposits, these events correspond unnecessarily to  one energetic $\gamma$-ray incidence, but with rather probability to multi $\gamma$-rays. For the events beyond 80 MeV, on the other hand, the spatial spread goes above $\delta_{\rm y}=3.5$ cm. These are mainly the cosmic $\gamma$-ray background recorded by random coincidence.

Fig. 1 (b) presents the spectrum of the total reconstructed energy. To understand the background, the beam-off spectrum is also recorded, as plotted by the red histogram after being normalized to the total live time of the beam experiment. It is clearly shown that, the high energy part above $E_{\rm tot}=80$ MeV is reproduced by the beam-off background. Assuming the CsI(Tl) crystals have the same responses to $\gamma$ and muon, the peak position is consistent with the Geant4 simulations \cite{Geant4,NIMA-CSHINE-Gamma}. In the region of $E_{\rm tot}<80$ MeV, on the other hand, the environmental $\gamma$ radiation is suppressed  by two orders of magnitude in comparison to the beam-on spectrum. In the following analysis, we merely concentrate on the 
regime of $E_{\rm tot}<80$ MeV.

\begin{figure} [h]
    \centering
    \includegraphics[width=0.43\textwidth]{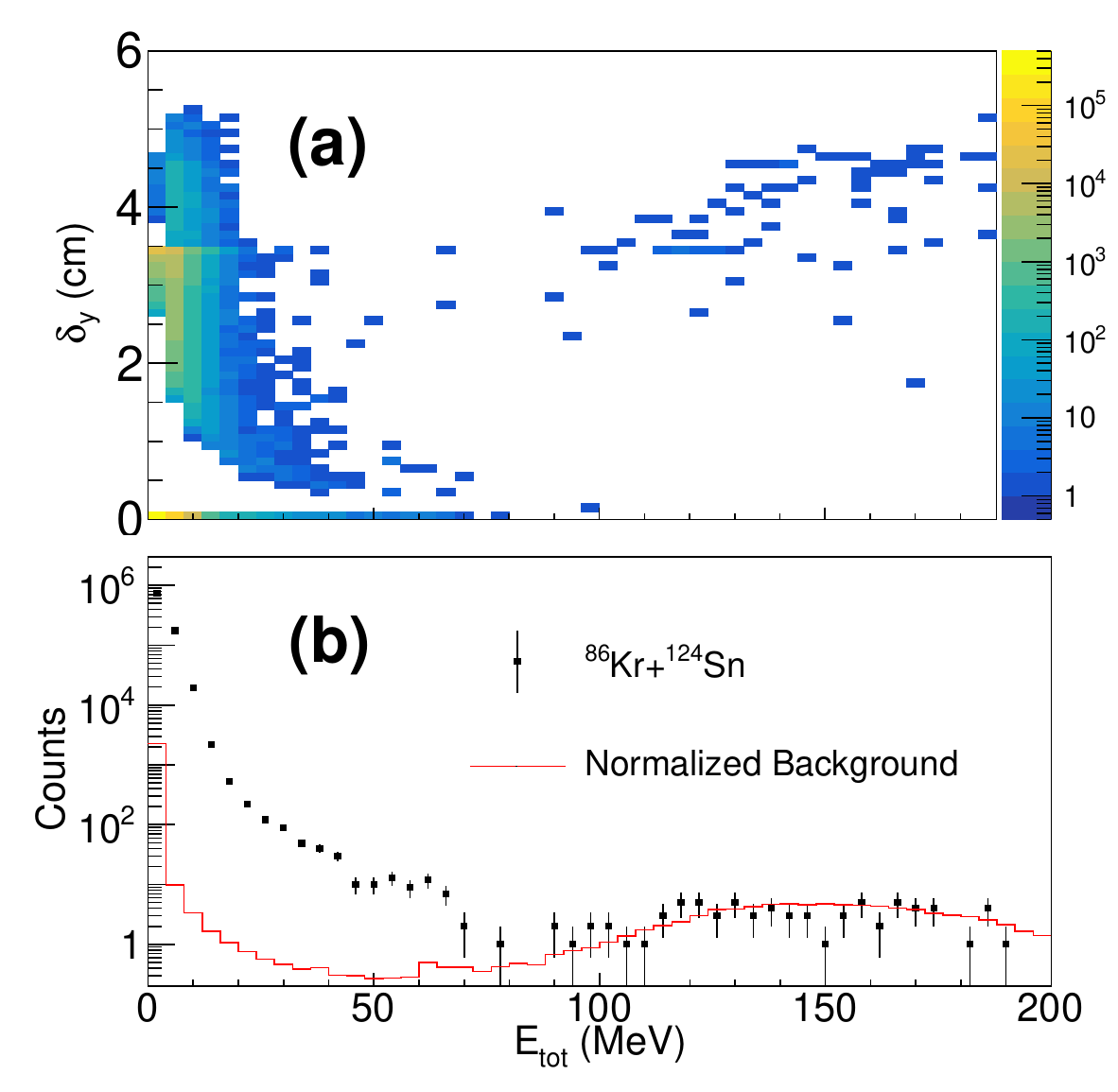}
    \caption{ (a) Vertical spatial spread $\delta_{\rm y}$ of the signals in the fired CsI(TL) units as a function of the reconstructed total energy, (b) The spectrum of the total reconstructed energy for the triggered events (solids squares)  and the beam-off background (histogram in red) normalized to equal live time of beam-on experiment. The error bars represent the statistic uncertainties only.}
    \label{fig:1}
\end{figure}

The IBUU transport model \cite{LI201829} is applied to simulate the collisions of \krsn~ by incorporating the effects of SRCs and HMT  \cite{xue16}. The behavior of HMT is quite similar for nuclei from deuteron to infinite nuclear matter, roughly exhibits a $C/k^{4}$ tail where $k$ is the single nucleon momentum \cite{Atti1996,Antonov1988}.  Here the IBUU model simulates collisions with typical initial HMT percentages, namely 0\% as free Fermi gas (FFG), 15\%, and 30\%. In principle, the bremsstrahlung photon production probability in reactions should be very small. Thus the influence of bremsstrahlung on the kinematics of nucleons is also small and a perturbative approach can be used to calculate the photon production probability at each $np$ collision. Based on the one boson exchange (OBE) approach, a good fitting expression of the elementary double differential photon production probability is applied in IBUU simulations \cite{gan94}

\begin{equation} \label {xsection}
d^2P/(d\Omega dE_{\gamma})=1.6\times10^{-7}[1-(E_{\gamma}/E_{\rm max})^2]^{\alpha}/(E_{\gamma}/E_{\rm max}),
\end{equation}
where $E_{\gamma}$ is energy of the produced photon and $E_{\rm max}$ is the total available energy in the CM system. The parameter $\alpha=0.7319-0.5898\beta_{i}$ and $\beta_{i}$ is the nucleon velocity  \cite{gan94}. The total photon production probability is then obtained by summing over all $np$ collisions over the entire history of the collision  \cite{gan94,xue16}.

\begin{figure}[h]
    \centering
    \includegraphics[width=.42\textwidth]{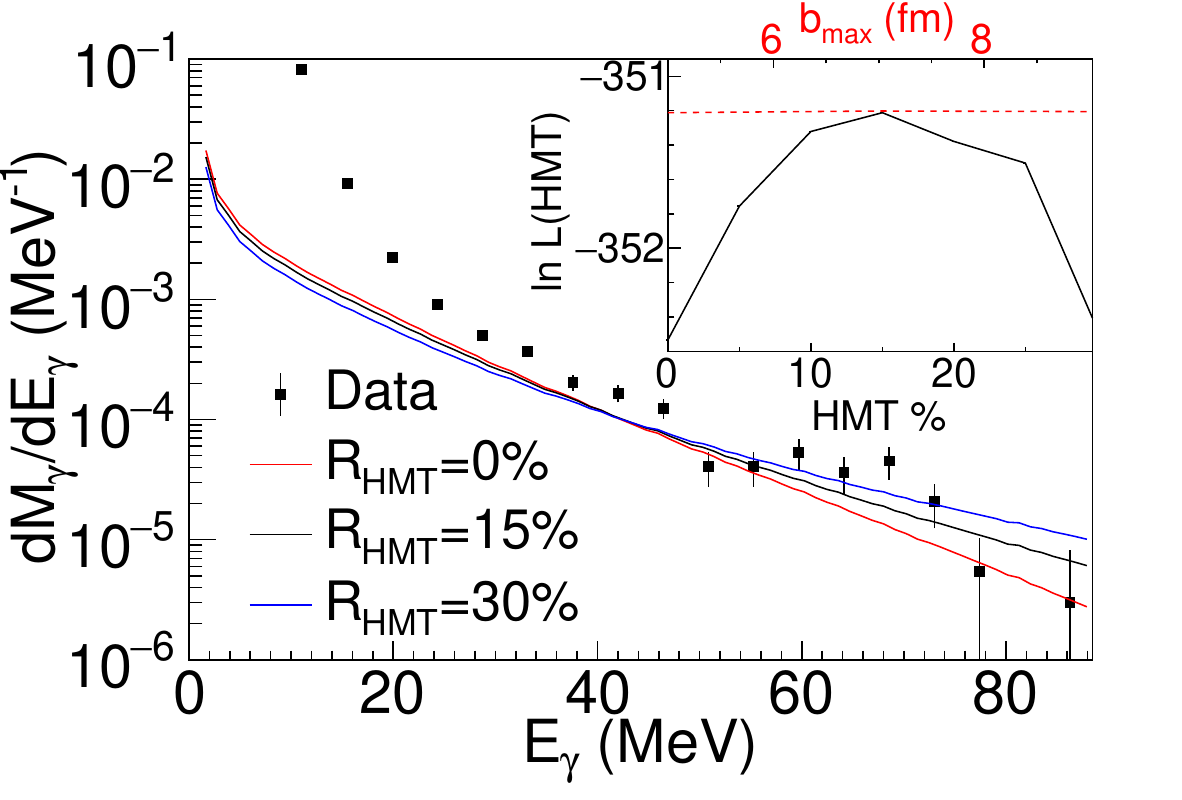}
    \caption{The energy spectrum of $\gamma$ rays in COM with the background being subtracted. Theoretical curves with different $R_{\rm HMT}$ by IBUU model calculation are all filtered by the detector responses obtained by Geant 4 simulation. No other normalization is introduced. The inset presents the likelihood distribution as a function of $R_{\rm HMT}$ (solid black) and $b_{\rm max}$ (dashed red), respectively. }
    \label{fig:2}
\end{figure}

{\it  Results and Discussions.}
Figure  \ref{fig:2} shows the $\gamma$ energy spectrum in CM reference after the background is subtracted. The ordinate is the differential multiplicity of the bremesstrahlung photons  obtained via 

\begin{equation} \label {calc}
\frac{dM_\gamma}{dE_\gamma} =\frac {Y(E_\gamma)}{N_{\rm 2B}}\frac{4\pi}{\Delta_{E_\gamma}\Omega_{\rm CsI}}
\end{equation}
where $Y(E_\gamma)$ is the $\gamma$ yield in the corresponding $E_\gamma$ bin and $N_{\rm 2B}$ is the total number of the triggered events with 2 FFs or 2 LCPs (or IMFs) detected. $\Delta_{E_\gamma}=4$ is the bin width and $\Omega_{\rm CsI}$ is the solid angle of the central CsI units in CM reference. The  saturation effect of the electronics is also corrected bin-by-bin  by full simulations of the detector response and the digitization process.  It is evident to observe the high energy component with $E_{\gamma}>30$ MeV, which exhibits  an exponential decrease with a slope very different from that in the region of $E_{\gamma}<20$ MeV, suggesting the bremsstrahlung radiation of high-energy  $\gamma$  is produced via the incoherent $np$ scattering process.


In order to confirm that the high-energy $\gamma$ rays are originated from the $np$ bremsstrahlung, the IBUU transport model simulations are performed by incorporating the process of $np\to np\gamma$ using formula (\ref{xsection}). The momentum-dependent potential parameter $x=-1$  \cite{LI201829} and the impact parameter range of $b\leq5$ fm are adopted in the simulation. The theoretic spectra are compared directly to the data after the  detector response filtering is done by Geant 4 simulation without introducing other normalization procedures, as shown by the curves in Fig. \ref{fig:2}.  It is shown that  the model predictions are in fair agreement with the experimental distribution for $E_\gamma>20$ MeV. 

By varying the HMT ratio $R_{\rm HMT}$, which is considered in sampling the initial momentum of the nucleons, the distribution starts to diverge in the high energy region of $E_{\gamma}>40$ MeV. In order to quantify the effect of HMT, one can perform the likelihood  analysis to find the most probable $R_{\rm HMT}$ giving the best description of the high energy $\gamma$ spectrum. The likelihood distribution varying with $R_{\rm HMT}$ is plotted as the solid lines in the inset. Here the range of analysis is set to $35 < E_\gamma < 80$ MeV.  With  $R_{\rm HMT}=15\%$, the calculated spectrum achieves the best description to the data points as given by the maximum likelihood value. It ends at a $\chi^2$ of about 2.64, equivalently  $90\%$ confidence to conclude there is an non-zero HMT contribution to the high-energy part of the $\gamma$ spectrum.  The low energy components of $E_{\gamma}<20$ MeV are dominated by collective resonance and statistic emissions, which are not included in the model for the current analysis. 

In IBUU simulations, the impact parameter $b$ brings the most influential uncertainty to the yield of bremsstrahlung photons, because the $np$ collision times varies significantly with the centrality. It is then necessary to check the effect of the uncertainty of $b$, although the trigger condition of containing 2 LCPs or 2 FFs has some selectivity of the collision geometry. We vary the maximum impact parameter $b_{\rm max}$ from 5 to 9 fm, the production probability of bremsstrahlung photons is changed by a factor of 1.7. But the likelihood keeps nearly constant as shown by the dashed line in the inset of Fig. \ref{fig:2}. It suggests that our conclusion favoring $15\%$ HMT is robust against the variation of $b_{\rm max}$.  The likelihood distribution changes negligibly with the N-N potential parameters in our simulation too.

More interestingly, apart from the agreement between the data and the model prediction, a hump-like structure following a dip at $E_{\gamma} \approx 50$ MeV is observed, which can reduce the confidence level of the likelihood analysis and lower slightly the HMT ratio at the best fit.  The deviation of the data points from  the expected curve near the dip is beyond the statistic uncertainty, thus it has suggestively other physical origins.

\begin{figure} [h]
    \centering
    \includegraphics[width=.42\textwidth]{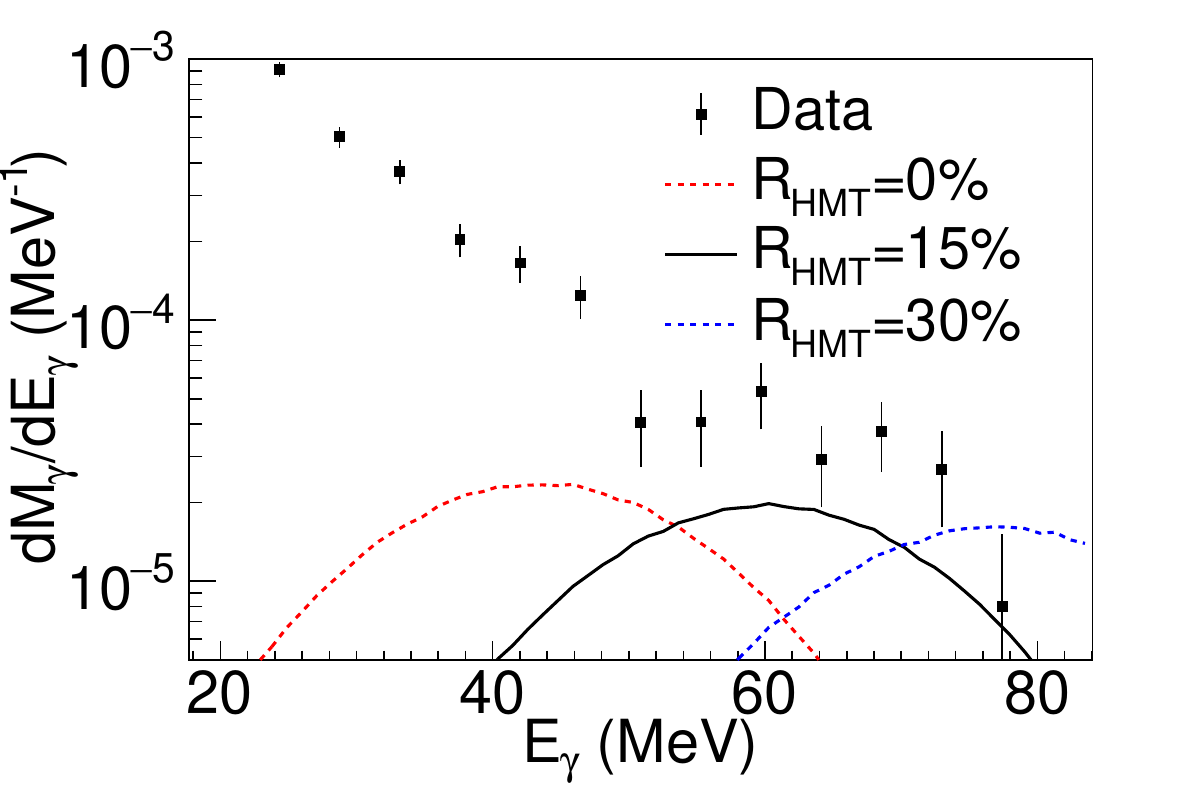}
    \caption{Model predictions  of the photon energy distribution from $np\to d\gamma$ channel with 0, 15\% and 30\% HMT ratio in the colliding nuclei in comparison to the experiment. }
    \label{fig:3}
\end{figure}

To further understand the dip and the broad hump-like structure on its right side, the capture process of  $np\to d\gamma$
is separately simulated by taking into account the HMT of nucleons too. Since it is a two-body final state for the capture process, the $\gamma$ is expected to be mono-energetic if the initial colliding energy of the $np$ pair in CM is fixed. In heavy ion collisions, the $\gamma$ energy distribution will be expanded due to the initial momentum variation.  To show the effect of the capture channel, the corresponding $\gamma$ spectra merely from $np\to d\gamma$ are plotted on top of the data in Fig. \ref{fig:3} with  $R_{\rm HMT}=0$, 15\% and 30\%, respectively.  As expected, the existence of HMT of nucleons enhances averagely the energy of the $np$ collision, and hence tends to move the wide energy peak of $np\to d\gamma$ rightwards. 
Although the absolute cross section of $np\to d\gamma$ is not determined in the simulations,  the peak position of distribution of $\gamma$ energies from  $np\to d\gamma$ is unique for a given ratio of HMT. The distribution with $R_{\rm HMT}=15\%$ agrees the best with the hump-like structure in the range of $45<E_{\gamma}<75$ MeV on the experimental spectrum, in accordance with the analysis in Fig. \ref{fig:2} where $R_{\rm HMT}15\%$ is favored. 
 
It is worth mentioning that in an independent experiment of $\rm {n+p \to n+p+\gamma}$, a dip-like structure is observed as well in the vicinity of $50$ MeV in reactions at 170 MeV beam energy \cite{Malek91,Sha91}, where  the $np\to d\gamma$ channel was later proposed  to account for the structure of the spectrum on a phenomenological base \cite{Brady95,Nif95}. 
All the results support that the capture channel $np\to d\gamma$ is at work in the production of bremsstrahlung $\gamma$-rays in heavy ion reactions. Further studies are required to elucidate its quantitative contribution and the possible interference with the $np \to np\gamma$ process.  

{\it  Conclusion.} The bremsstrahlung $\gamma$ rays are measured in the reaction of \krsn~ at 25 MeV/u beam energy. The $\gamma$ energy spectrum above 30 MeV were analyzed in an IBUU model incorporating the bremsstralung photon production from $np$ incoherent scattering process. The HMT of the nucleons in the colliding nuclei is included by varying the ratio. It is demonstrated that the yield of high energy $\gamma$-rays are sensitive to the HMT of nucleons in nuclei, with a 15\% ratio being favored by the analysis. Moreover, a dip structure is observed in the $\gamma$ spectrum, in accordance with some previous independent experiments, suggesting that the $np\to d\gamma$  channel is at work. The experiment proves that the bremsstrahlung $\gamma$ production is a new means to study the SRC in atomic nuclei. Further systematic measurements with more statistics extending to higher energy of the $\gamma$ rays are particularly favored.  
 
{\it  Acknowledgements.}
 This work is supported  by the Ministry of Science and Technology of China under Grant Nos. 2022YFE0103400 and 2020YFE0202001,
 by the National Natural Science Foundation of China under Grant Nos. 11961141004, 12275129  
 and by Tsinghua University Initiative Scientific Research Program and the Heavy Ion Research Facility at Lanzhou (HIRFL).

\end{document}